\pdfoutput=1
\documentclass[aps,prd,amsmath,floats,floatfix,twocolumn,
superscriptaddress,nofootinbib,showpacs]{revtex4-1}

\usepackage[T1]{fontenc}
\usepackage[utf8]{inputenc}
\usepackage{lmodern}
\usepackage{verbatim}

\usepackage[dvipsnames]{xcolor}
\definecolor{linkcolor}{rgb}{0.0,0.3,0.5}
\usepackage[hypertexnames=false, unicode, colorlinks=true, linkcolor=linkcolor,
citecolor=linkcolor, filecolor=linkcolor,urlcolor=linkcolor,
pdfusetitle]{hyperref}

\usepackage[all]{hypcap}
\usepackage{graphicx}
\usepackage{xspace}
\usepackage{amssymb}
\usepackage[normalem]{ulem} 
\usepackage{bm} 
\usepackage{enumitem,amssymb}
\usepackage{microtype}
\usepackage[english]{babel}
\usepackage{blindtext}

\begin{document}

\title{A Tale of Two Calibrators: Comparing Newtonian and Photon Calibration in Advanced LIGO after the Third Observing Run}

\author{Dana Jones}
\email{dana.jones@anu.edu.au}
\affiliation{OzGrav-ANU, Centre for Gravitational Astrophysics, Research School of Physics and Research School of Astronomy \& Astrophysics, The Australian National University, ACT 2601, Australia}
\affiliation{Center for Experimental Nuclear Physics and Astrophysics, University of Washington, Seattle, WA 98195, USA}

\author{Michael P. Ross}
\affiliation{Center for Experimental Nuclear Physics and Astrophysics, University of Washington, Seattle, WA 98195, USA}

\author{Ling Sun}
\affiliation{OzGrav-ANU, Centre for Gravitational Astrophysics, Research School of Physics and Research School of Astronomy \& Astrophysics, The Australian National University, ACT 2601, Australia}

\author{Jeffrey S. Kissel}
\affiliation{LIGO Hanford Observatory, Richland, WA 99352, USA}

\author{Dripta Bhattacharjee}
\affiliation{LIGO Livingston Observatory, Livingston, LA 70754, USA}
\affiliation{Institute of Multi-messenger Astrophysics and Cosmology, Missouri Institute of Science and Technology, Rolla, MO 65409, USA}
\affiliation{Kenyon College, Gambier, OH 43022, USA}

\author{Anthony Sanchez}
\affiliation{LIGO Hanford Observatory, Richland, WA 99352, USA}

\author{Bram J.J. Slagmolen}
\affiliation{OzGrav-ANU, Centre for Gravitational Astrophysics, Research School of Physics and Research School of Astronomy \& Astrophysics, The Australian National University, ACT 2601, Australia}

\author{Jens Gundlach}
\affiliation{Center for Experimental Nuclear Physics and Astrophysics, University of Washington, Seattle, WA 98195, USA}

\hypersetup{pdfauthor={Jones et al.}}

\date{\today}


\begin{abstract}

Precise calibration of LIGO's strain readout plays a vital role in our ability to extract information from gravitational-wave detections.
We present a comparative analysis of two independent direct-force calibration methods: the Newtonian Calibrator (NCal) and Photon Calibrator (PCal) systems. Currently, LIGO relies solely on the PCal system as its absolute calibration reference.
To facilitate comparison, a series of direct-force injections were performed using the NCal and PCal systems at the LIGO Hanford observatory just after the completion of the third observing run. We compute the ratio of the measured to expected strain amplitudes for every injection. For each injection frequency across a 30~Hz band, the ratios of NCal and PCal differ by $\sim$0.5--1\%, demonstrating a small systematic difference between the two systems. This offset highlights the importance of maintaining multiple independent calibration references to validate LIGO's calibration.

\end{abstract}


\maketitle


\section{Introduction}
\label{sec:introduction}

Over the past decade, several hundred gravitational-wave signals have been detected by the LIGO--Virgo--KAGRA (LVK) Collaboration~\cite{aLIGO, aVirgo, KAGRA}. These observations have provided unprecedented insight into the properties and formation channels of compact object binary systems, while also enabling new tests of general relativity in the strong-field regime and independent measurements of cosmological parameters~\cite{GWTC-4, GWTC-5}. However, accurate inference of source parameters relies critically on the precise and accurate reconstruction of the strain data measured by the observatories~\cite{Sun2020}. Strain is a dimensionless quantity equal to the differential length change of the detector arms relative to their original length.

A gravitational-wave interferometer operates as a closed-loop feedback control system that actively suppresses differential arm-length changes induced by external disturbances up to $\sim 300$~Hz. Thus, the gravitational-wave strain cannot be measured directly from the interferometer output. Instead, it must be reconstructed from the detector response using the digitized error and control signals recorded by the interferometer's sensing and actuation systems. Detector calibration is the process of converting the raw detector output into a calibrated strain time series through a detector response model~\cite{GW150914_calibration, Sun2020, Wade2025}.
The response model, which describes how differential arm-length displacement is sensed, controlled, and actuated within the interferometer, consists of time- and frequency-dependent sensing, digital control, and actuation functions. Together, they determine the mapping between the physical differential arm motion of the detector and the reconstructed strain. Accurate calibration therefore requires precise characterization of each stage of the detector response.

Achieving this characterization is challenging because gravitational-wave detectors are complex instruments with numerous noise sources, imperfectly known transfer functions, and parameters that drift over time. To limit systematic error and uncertainty in the calibrated strain, we must measure several quantities with high accuracy and precision: the optical power circulating in the two Fabry-Perot arm cavities, the masses and suspension dynamics of the test masses, the detector timing and frequency-reference systems, and the electronic gains that convert between digital signals, voltages, and currents~\cite{Wade2025}. Uncertainties in all of these quantities propagate through the detector response model and ultimately into the calibrated strain data.

Although calibration errors have not historically had a major impact on the detection of gravitational-wave signals, they are expected to become a limiting factor in source parameter estimation for next-generation detectors~\cite{Sun2020, Evans2021}. More broadly, calibration systematics can bias astrophysical population studies, measurements of cosmological parameters, and tests of general relativity~\cite{Wade2025}.
To mitigate these effects, several independent direct-force absolute calibration references have been developed. They can be used to apply forces with known, standards-traceable magnitudes directly to the test masses, providing independent references against which the detector response model can be validated.
However, there is no such thing as a ``perfect'' standard; even these absolute references have associated uncertainties which in turn contribute to the overall uncertainty in the reconstructed detector strain.

In the Advanced LIGO, Virgo (prior to the LVK's fourth observing run [O4]), and KAGRA detectors, Photon Calibration (PCal) acts as the primary direct-force absolute calibration system~\cite{aLIGO, aVirgo, KAGRA, Karki2016, Estevez2021_1, Karki_thesis, Cahillane2017, Bhattacharjee2020, Sun2020, Sun2021, Wade2025, Chen2025}. The PCal system comprises a series of auxiliary lasers that apply a known force to the test masses via modulated radiation pressure. The reflected light is detected with photodiodes that have been calibrated against standards from the National Institute of Standards and Technology (NIST) and the Physikalisch-Technische Bundesanstalt (PTB), which enable precise determination of the force acting upon the test masses~\cite{Bhattacharjee2024}.
An alternative calibration technique, Newtonian Calibration (NCal)~\cite{Hirakawa1980, Kuroda1985, Mio1987}, has been tested and/or implemented in the LIGO~\cite{Ross2021}, Virgo~\cite{Estevez2018, Estevez2021_2, Aubin2024}, and KAGRA~\cite{Inoue2018} detectors, and it has acted as Virgo's primary absolute calibration system since the beginning of O4~\cite{Virgo_O4_cal}. The NCal system employs a rotor (or series of rotors~\cite{Estevez2021_2, Ross2023}) with an asymmetric mass distribution that, when rotated, imparts a time-varying gravitational force on the test mass.

In addition to direct-force absolute calibration references, several indirect approaches exist.
These include applying a frequency modulation to the primary laser, which produces a well-defined effective modulation of the arm length~\cite{Goetz2010_1}, using auxiliary lasers to apply a known radiation-pressure force to the test masses~\cite{GW150914_calibration}, and making measurements based directly on the laser wavelength with the detector temporarily configured as a free-swinging Michelson interferometer, such that the observed interference fringes provide an absolute length reference~\cite{GW150914_calibration, Abadie2010, Accadia2011}.
Another approach, known as astrophysical calibration, involves using gravitational-wave sources themselves as calibration standards~\cite{Essick2019, Schutz2020, astro_calibration}. However, the precision achieved by direct-force calibration techniques well exceeds that of these indirect techniques~\cite{GW150914_calibration, Goetz2010_2}.

Incorporating two independent absolute references such as the NCal and PCal systems within a single detector is highly beneficial, as it not only enables cross-validation, but also ensures accurate calibration across the full detector bandwidth.
To this end, an NCal rotor was installed in the LIGO Hanford observatory during the third observing run and then spun at a series of frequencies. This imparted a time-varying force on the test mass in the $x$-arm~\cite{Ross2021}. Meanwhile, the PCal system was used to inject a force on the test mass in the $y$-arm at a set of similar frequencies to enable direct comparison.
In this paper, we analyze this set of injections to quantify and compare the ratio of measured to expected strain amplitudes across the NCal and PCal systems. We also consider how these two calibration methods can be combined to improve the overall accuracy and reduce the error and uncertainty associated with detector calibration.

The paper is organized as follows:
In Sec.~\ref{sec:NCal}, we give an overview of the concept of NCal. In Sec.~\ref{sec:data}, we introduce the data used in this study to compare the NCal and PCal systems in LIGO. Section~\ref{sec:analysis} provides details on the data analysis procedure, while in Sec.~\ref{sec:results}, we discuss the results of this study and consider their implications. We conclude in Sec.~\ref{sec:conclusion}. In Appendix~\ref{appendix:error_prop}, we derive the full error propagation.

This work builds upon Ref.~\cite{Ross2021}, in which LIGO's NCal was first introduced. Hence, we refer the reader to Ref.~\cite{Ross2021} for a detailed description of the NCal design and initial experimental characterization.


\section{Newtonian calibration}
\label{sec:NCal}

Gravitational calibration is characterized by an oscillating mass distribution (e.g., a non-uniform spinning rotor) that produces time-varying changes in the local gravitational field. For a nearby test mass, this induces a time-dependent force, driving the test mass by a well-defined displacement. In the context of gravitational-wave interferometers, a gravitational calibrator positioned near a test mass in one arm will induce a periodic differential arm-length displacement that is indistinguishable from the detector response to a passing gravitational wave.

\begin{figure*}[hbt!]
    \centering
    \includegraphics[trim = 4.5cm 1cm 4.5cm 1cm, clip, width=0.7\linewidth]{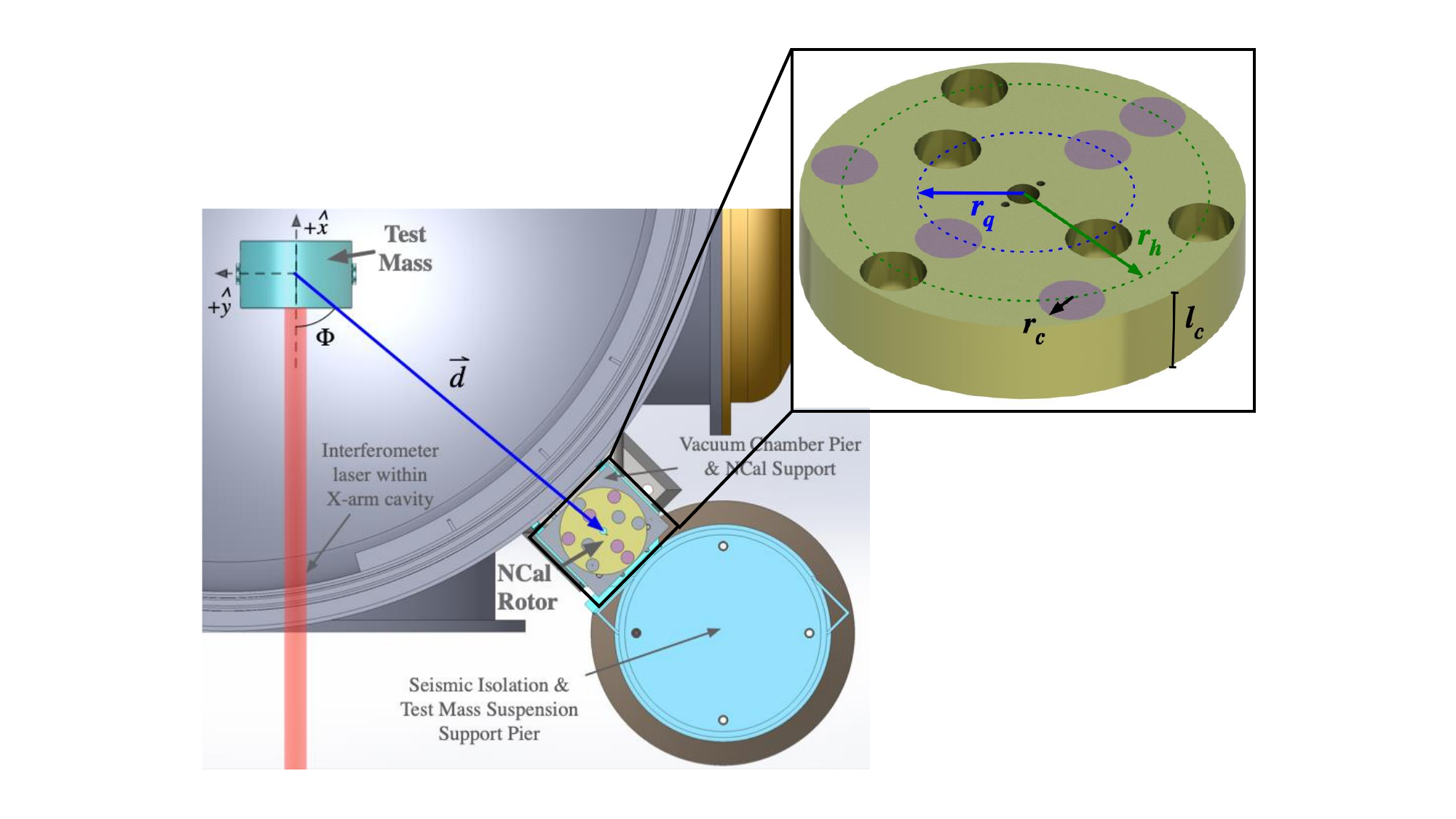}
    \caption{Simplified CAD rendering of the NCal and test mass system as installed in the LIGO Hanford observatory. The box in the top right corner shows the NCal rotor in more detail, where the purple circles represent tungsten cylinders embedded in the rotor and the gold circles represent the cylindrical holes in the rotor that remain unfilled. (Figure adapted from Ref.~\cite{Ross2021}.)}
    \label{fig:ncal}
\end{figure*}

Gravitational calibration operates purely within the Newtonian limit, wherein a gravitational force is simply a superposition of forces created by each mass multipole moment. Hence, a rotor will exert a gravitational force on the test mass at any given multiple of its rotational frequency $f$ when it contains a sub-geometry that is symmetric at that same multiple. Meanwhile, any sub-geometry whose mass distribution does not change as the rotor spins will not contribute to the total force.
As an example, the rotor used in this study (pictured in Fig.~\ref{fig:ncal}) was designed to have 2-fold and 3-fold symmetric mass distributions, so it induces a force on the test mass at both $2f$ and $3f$. See Ref.~\cite{Ross2021} for a more detailed breakdown of the NCal system, including the rotor components, the drive system used to spin the rotor, and the test mass upon which the rotor exerts a force.

While the magnitude of the force is independent of the rotor's spin frequency, it is increasingly suppressed at higher-order multipoles (due to their more rapid spatial decay). In other words, the force induced by a two-fold symmetric quadrupole mass distribution scales as $d^{-4}$, while for a three-fold symmetric hexapole system, it scales as $d^{-5}$.

In Ref.~\cite{Ross2021}, the authors established NCal as a promising independent method for the absolute calibration of LIGO. They designed and installed an NCal in the LIGO Hanford observatory and then spun it up incrementally, injecting a signal into the detector at various frequencies. The authors found good agreement between the detector’s reconstructed signal and the model predictions, with a relative uncertainty of $<1$\% in signal amplitude for injected frequencies above approximately 10~Hz. However, the relatively short duration of the injections limited the precision of the measurements, resulting in statistical uncertainties that exceeded the model uncertainties. The authors also note that a combined analysis of NCal and PCal injections would provide a more precise absolute calibration. In this work, we address these limitations through longer-duration, higher-sensitivity data runs in which NCal and PCal signals have been injected simultaneously, enabling a direct comparison between the two calibration systems.


\section{The data}
\label{sec:data}

To assess the consistency between NCal and PCal, we use a dataset in which injections were performed at a common set of frequencies with each calibration system. The resulting strain amplitudes could then be compared at each discrete frequency, providing a direct measure of the broadband agreement between the two calibration techniques. The data collection is described below.

\begin{figure*}[hbt!]
    \centering
    \includegraphics[width=1.0\linewidth]{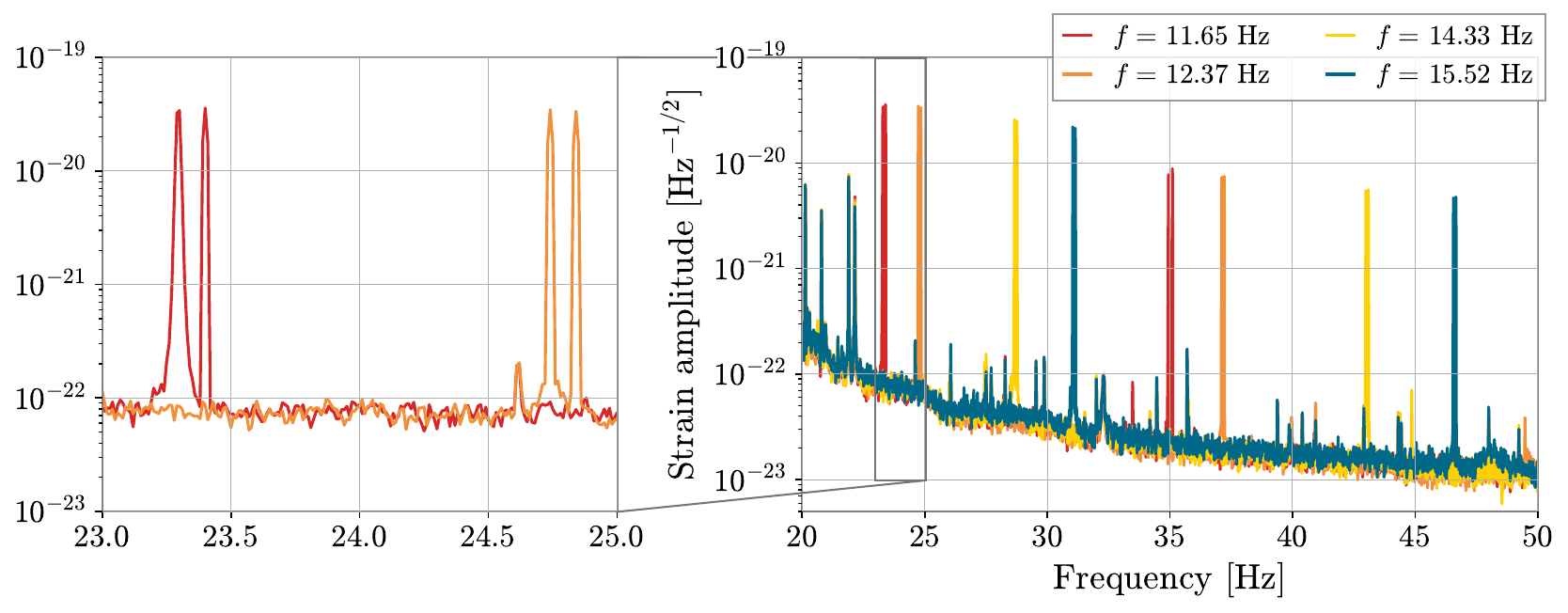}
    \caption{Amplitude spectral density of the Hanford observatory's strain output computed over the time periods corresponding to each pair of NCal and PCal injections. Only the $2f$ and $3f$ contributions, which dominate for our rotor configuration, are shown. The rotor spin frequency $f$ corresponding to each NCal injection is indicated by the line color. Each NCal and PCal frequency-pair are separated by approximately 0.1~Hz (inset), where the larger of the two corresponds to the NCal injection (see odd rows in Table~\ref{tab:NCal_PCal_freqs}). The injections for which the NCal and PCal frequencies are swapped (see even rows in Table~\ref{tab:NCal_PCal_freqs}) are not pictured in this figure but are visually nearly identical to the set shown here. The frequency resolution of this spectrum is 0.01~Hz.}
    \label{fig:asd}
\end{figure*}

On September 3rd, 2020, LIGO Hanford's NCal rotor was spun at four different frequencies sequentially between 11 Hz and 16 Hz. Because of the inbuilt quadrupolar and hexapolar mass distributions, this injected a gravitational force into the detector at twice ($2f$) and thrice ($3f$) the spin frequency ($f$).
The PCal system was run simultaneously in the $y$-arm, forcing the corresponding test mass at two frequencies such that, for every $2f$ and $3f$ frequency pair injected by NCal, there was a corresponding PCal frequency pair offset from the NCal frequencies by $+0.1$~Hz.
To account for any potential frequency dependence in the detector's response to test mass displacement, an additional set of injections was performed in which the NCal and PCal frequencies were swapped.
For each set of injections, we consider a 30-minute segment of data starting after the NCal and PCal systems have settled for a few minutes at their respective frequencies.
To summarize, at any given moment, four signals were being injected at two sets of frequencies---each set comprising one NCal and one PCal signal spaced $0.1$ Hz apart. We show the amplitude spectral density (ASD) of each set of injections in Fig.~\ref{fig:asd}.

We extracted the precise NCal rotor spin frequencies and their harmonics used in the above injections from the rotor's angle encoder channel.\footnote{The exact channel name is \texttt{H1:CAL-NCALX\_ENCODER\_VELOCITY \_OUT\_DQ}.} These frequencies are shown in Table~\ref{tab:NCal_PCal_freqs}, with an associated uncertainty of $\sigma_f = 5 \times 10^{-5}$~Hz. Table~\ref{tab:NCal_PCal_freqs} also includes the frequencies corresponding to the associated PCal injections and the start and end GPS times for each data run.
We used two additional data channels in this analysis: 1) the PCal reflected photodiode channel,\footnote{The exact channel name is \texttt{H1:CAL-PCALY\_RX\_PD\_OUT\_DQ}.} from which we calculated the expected strain amplitude of the PCal injections, and 2) the calibrated strain channel,\footnote{The exact channel name is \texttt{H1:GDS-CALIB\_STRAIN}.} from which we extracted the strain of both the NCal and PCal injections as measured by LIGO Hanford.

\begin{table*}[tbh]
    \centering
    \setlength{\tabcolsep}{7pt}
    \renewcommand\arraystretch{1.2}
    \begin{tabular}{lccccccc}
        \hline
        \hline
         & NCal ($f$) [Hz] & NCal ($2f$) [Hz] & NCal ($3f$) [Hz] & PCal ($2f$) [Hz] & PCal ($3f$) [Hz] & $t_{\rm start}$ [s] & $t_{\rm end}$ [s] \\
        \hline
        1 & 11.6477 & 23.2954 & 34.9432 & 23.4000 & 35.1000 & 1283182218 & 1283184018 \\
        2 & 11.7000 & 23.4000 & 35.1000 & 23.2969 & 34.9453 & 1283184318 & 1283186118 \\
        3 & 12.3700 & 24.7400 & 37.1100 & 24.8400 & 37.2100 & 1283186718 & 1283188518 \\
        4 & 12.4200 & 24.8400 & 37.2600 & 24.7422 & 37.1094 & 1283188818 & 1283190618 \\
        5 & 14.3300 & 28.6599 & 42.9899 & 28.7600 & 43.0900 & 1283190918 & 1283192718 \\
        6 & 14.3800 & 28.7600 & 43.1399 & 28.6641 & 42.9922 & 1283193018 & 1283194818 \\
        7 & 15.5200 & 31.0399 & 46.5599 & 31.1400 & 46.6600 & 1283195178 & 1283196978 \\
        8 & 15.5700 & 31.1400 & 46.7100 & 31.0391 & 46.5625 & 1283197278 & 1283199078 \\
        \hline
        \hline
    \end{tabular}
    \caption{NCal rotor spin frequency ($f$), the harmonics ($2f$ and $3f$) at which a gravitational force is injected into the detector, the corresponding PCal injection frequencies ($2f$ and $3f$), and the span of time (in GPS time units) over which the data is valid for each set of injections.}
    \label{tab:NCal_PCal_freqs}
\end{table*}

In the following analysis, we have considered only the magnitude of the NCal and PCal forces. The NCal system contains two rotary encoders that independently monitor the rotation frequency, one on the top and one on the bottom of the shaft that runs through the center of the rotor. While these encoders should be able to measure the rotor phase, the absolute phase between the encoders and the rotor masses was not properly aligned during installation of the NCal system at Hanford, rendering us unable to predict the phase of the NCal forces.


\section{Analysis}
\label{sec:analysis}


In this section, we describe the data analysis procedure that was designed to facilitate direct comparison between the LIGO detectors' current absolute calibration reference, PCal, and the direct-force reference NCal. In particular, we prepare all the necessary ingredients to calculate the ratios of the measured to expected strain amplitudes for each NCal and PCal injection, i.e., $R = h_{\rm meas} / h_{\rm exp}$. To quantify $R$ across all injections, we begin by cleaning the time series data to remove glitches (Sec.~\ref{sec:chisqr}). Then we extract the measured strain amplitudes of the NCal and PCal injections (Sec.~\ref{sec:meas_strain}), quantify the expected strain amplitudes (Sec.~\ref{sec:exp_strain}), and apply a torque correction to the NCal strain measurements (Sec.~\ref{sec:torque_corr}).

\subsection{$\chi^2$ cuts}
\label{sec:chisqr}

While the noise in gravitational-wave interferometers is generally thought to be Gaussian and stationary, non-Gaussian, short-duration artifacts known as glitches can occasionally appear~\cite{Soni2025}. The time series data analyzed in this study was contaminated by several broadband glitches, so we cleaned the data before analyzing. We describe this process below, wherein we used a $\chi^2$-test to remove outliers.

We started with a segment of time series data $x$. The $\chi^2$-statistic of $x$ can be written as
\begin{equation}
\chi^2 = \sum_{i=1}^{n} \frac{(x_i - \bar{x})^2}{s^2}.
\end{equation}
We first whitened $x$ such that the mean and variance above became $\bar{x} \approx 0$ and $s^2 \approx 1$. Then, the square of the whitened data could be described by a $\chi^2$ distribution with one degree of freedom: $x^2 \sim \chi^2(1)$.
We empirically chose a $\chi^2(1)$ threshold of 5.5 to identify large excursions in the whitened strain. This threshold corresponds to approximately the 98th percentile of the theoretical $\chi^2(1)$ distribution, implying a $\sim 2$\% false alarm probability per sample.
We treated all $x_i^2$ data points above this threshold as outliers and removed them from the dataset before interpolating across the remaining data to fill in the gaps that had been left behind.


\subsection{Measured strain}
\label{sec:meas_strain}

In this section, we explain how we extracted the amplitude of the measured strain corresponding to each NCal and PCal injection. We used a process called \textit{demodulation}, which involves heterodyning the data, averaging the oscillating part over a short duration known typically as the \textit{stride}, and then extracting the magnitude $A$ and phase $\phi$ of the slowly varying complex amplitude that remains.

We demodulated the calibrated strain time series at each relevant frequency for the corresponding segment of time using the \texttt{demodulate} method from the \texttt{gwpy.Timeseries} library~\cite{gwpy}.
We chose a stride $\approx 200$~s such that the algorithm made nine consecutive measurements of the amplitude, generating a distribution with $N = 9$ data points. However, the precise stride was chosen to be an integer multiple of the period of each injected signal so that the demodulation window consistently aligned with the signal peak. 
We then took an average of the $N$ consecutive measurements to obtain a final value for the strain magnitude:
\begin{equation}
    h^{\rm meas}_{\rm NCal} = \frac{\sum_{i=1}^{N} A_{i, {\rm NCal}}^{\rm meas}}{N}, \quad h^{\rm meas}_{\rm PCal} = \frac{\sum_{i=1}^{N} A_{i, {\rm PCal}}^{\rm meas}}{N},
    \label{eqn:meas_strain}
\end{equation}
where $A_{i, {\rm NCal}}^{\rm meas}$ and $A_{i, {\rm PCal}}^{\rm meas}$ are the measured strain amplitudes of each NCal and PCal injection, respectively.
The associated uncertainties on these measurements are
\begin{equation}
    \sigma_{h, \rm{NCal}}^{\rm meas} = \frac{s_{h, \rm NCal}^{\rm meas}}{\sqrt{N}}, \quad \sigma_{h, \rm{PCal}}^{\rm meas} = \frac{s_{h, \rm PCal}^{\rm meas}}{\sqrt{N}},
    \label{eqn:meas_strain_unc}
\end{equation}
where $s_{h, \rm NCal}^{\rm meas}$ and $s_{h, \rm PCal}^{\rm meas}$ are, respectively, the standard deviations of the sets of measurements \{$A_{\rm NCal}^{\rm meas}$\} and \{$A_{\rm PCal}^{\rm meas}$\}.


\subsection{Expected strain}
\label{sec:exp_strain}

In this section we describe how we calculated the expected strain amplitude of each NCal and PCal injection. This was done analytically for the NCal injections and empirically for the PCal injections (using the appropriate data channel).

We estimated the expected strain amplitude produced by each NCal injection using the following equation~\cite{Ross2021}:
\begin{equation}
    h^{\rm exp}_{\rm NCal} = \frac{F^{\rm exp}_{\rm NCal}}{M (2 \pi f)^2 L_x},
    \label{eqn:NCal_exp_strain}
\end{equation}
where $F^{\rm exp}_{\rm NCal}$ is the expected gravitational force induced by the NCal rotor on the test mass, $M$ is the mass of the test mass, and $L_x$ is the $x$-arm length (see Table~\ref{tab:NCal_params}). Because the force amplitude for a fixed NCal geometry is independent of the rotor's spin frequency, we only needed one value of $F^{\rm exp}_{\rm NCal}$ for each frequency harmonic, in this case $2f$ and $3f$. These two forces were determined numerically in Ref.~\cite{Ross2021} and are also shown along with their uncertainties in Table~\ref{tab:NCal_params}.

\begin{table*}[tbh]
    \centering
    \setlength{\tabcolsep}{15pt}
    \renewcommand\arraystretch{1.2}
    \begin{tabular}{lllcc}
        \hline
        \hline
        Parameter & Symbol & Unit & Mean & Uncertainty \\
        \hline
        Test mass & $M$ & kg & 39.68 & $10^{-3}$ \\
        $x$-arm length & $L_x$ & m & 3999.49 & 0.003 \\
        $y$-arm length & $L_y$ & m & 3999.46 & 0.003 \\
        Expected force\,($2f$) & $F^{\rm exp}_{\rm NCal}\,(2f)$ & pN & 19.16 & 0.14 \\
        Expected force\,($3f$) & $F^{\rm exp}_{\rm NCal}\,(3f)$ & pN & 9.06 & 0.09 \\
        Torque force ($2f$) & $F^{\rm torq}_{\rm NCal}\,(2f)$ & pN & 0.339 & 0.026 \\
        Torque force ($3f$) & $F^{\rm torq}_{\rm NCal}\,(3f)$ & pN & 0.199 & 0.015 \\
        Torque phase\,($2f$) & $\phi\,(2f)$ & rad & 0.469 & -- \\
        Torque phase\,($3f$) & $\phi\,(3f)$ & rad & $-0.553$ & -- \\
        \hline
        \hline
    \end{tabular}
    \caption{Various parameters and their uncertainties used in the NCal and PCal analyses.}
    \label{tab:NCal_params}
\end{table*}

To estimate the expected strain amplitude produced by each PCal injection, we used the PCal reflected photodiode channel. Similar to the measured strain calculations, we demodulated the time series at the relevant frequencies. However, because the PCal channel is significantly more noisy than the calibrated strain channel, we first applied a finite impulse response band-pass filter with a band of $\pm 20$~Hz centered on each frequency of interest, and then we demodulated the filtered time series with the same strides as above. Again averaging these successive amplitude measurements, we found
\begin{equation}
    A^{\rm exp}_{\rm PCal} = \frac{\sum_{i=1}^{N} A_{i, {\rm PCal}}^{\rm exp}}{N}, \quad \sigma_{A, \rm PCal}^{\rm exp} = \frac{s_{A, \rm PCal}^{\rm exp}}{\sqrt{N}}.
    \label{eqn:PCal_exp_unc}
\end{equation}
Then, because the time series was not in strain units, we needed to divide by $L_y$ and apply an additional correction factor $C$---to convert the amplitude into strain and undo any filters that were applied---as shown:
\begin{equation}
    h^{\rm exp}_{\rm PCal} = C \frac{A^{\rm exp}_{\rm PCal}}{L_y}.
    \label{eqn:PCal_exp_strain}
\end{equation}
We found $C$ using \texttt{compute\_pcal\_correction} from the \texttt{pydarm.pcal} module~\cite{pydarm}.


\subsection{Torque correction}
\label{sec:torque_corr}

\begin{figure}[hbt!]
    \centering
    \includegraphics[trim = 5.85cm 0cm 5.5cm 0cm, clip, width=1.0\linewidth]{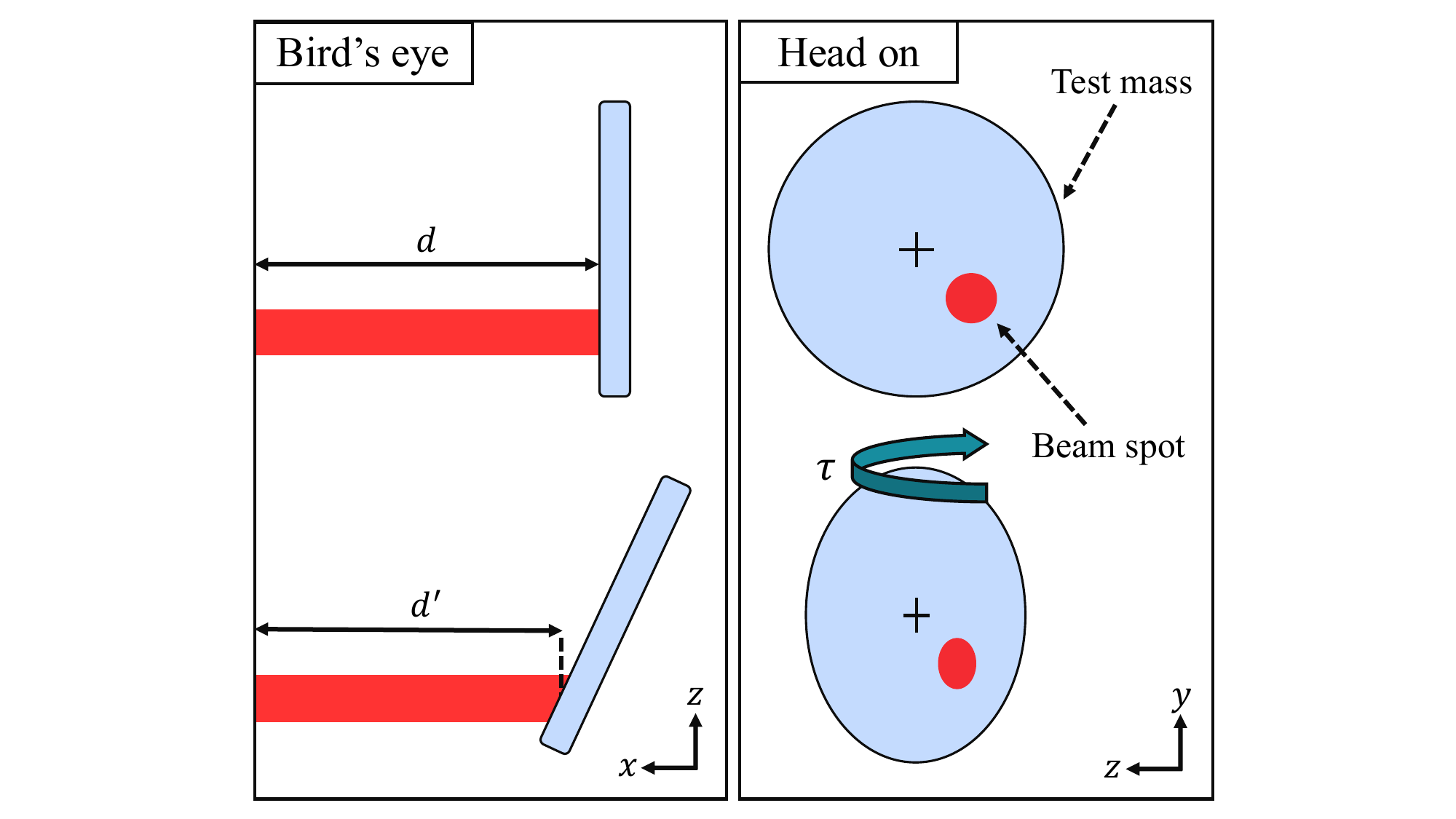}
    \caption{Simplified graphic showing head on (right panel) and bird's eye (left panel) views of the interferometer test mass (light blue) and the main laser (red), which is off-center. When the NCal rotor (not shown) applies a torque to the test mass (blue arrow), it causes a change in the path length of the laser, as shown on the bottom left. If not accounted for, this torque couples into the strain measurement, biasing the calibration. (Sizes and offsets are not to scale and are exaggerated for clarity.)}
    \label{fig:test_mass_torque}
\end{figure}

Because the NCal rotor was not installed in-line with the test mass, in addition to a force, it also applies a torque on the test mass. For a configuration in which the interferometer beam is perfectly centered on the test mass, this torque would not couple into the strain measurements. However, this was not the case when this dataset was taken.
Due to mirror imperfections at the center of the test mass, the main laser beam was offset from the center of rotation in both the $y$- and $z$-directions~\cite{Sun2021, Brooks2021}. Hence, when the NCal rotor was spun, it induced rotations in the test mass about the $y$- and $z$-axes (known as \textit{pitch} and \textit{yaw}). This in turn induced a change in the path length of the laser beam (see Fig.~\ref{fig:test_mass_torque} for a basic illustration). If left uncorrected, this shifted path length would result in an incorrect force reading.

We started with the uncorrected measured force, which scales with the measured strain found in Sec.~\ref{sec:meas_strain}, as
\begin{equation}
    F^{\rm meas}_{\rm NCal} = h^{\rm meas}_{\rm NCal}\, M (2 \pi f)^2 L_x.
    \label{eqn:NCal_F_meas}
\end{equation}
Using the torque to force coupling explored in Ref.~\cite{Ross2021}, we applied an apparent force correction, $F^{\rm torq}_{\rm NCal}$, to the measured force as shown:
\begin{equation}
    F^{\rm meas \,\prime}_{\rm NCal} = \sqrt{F^{\rm torq \, 2}_{\rm NCal} + F^{\rm meas \, 2}_{\rm NCal} - 2 \big(F^{\rm torq}_{\rm NCal} F^{\rm meas}_{\rm NCal}\big)^2 \cos(\phi)},
    \label{eqn:NCal_F_corr}
\end{equation}
where $\phi$ is the relative phase between $F^{\rm torq}_{\rm NCal}$ and $F^{\rm meas}_{\rm NCal}$. (The values of $F^{\rm torq}_{\rm NCal}$ and $\phi$ for the $2f$ and $3f$ harmonics are shown in Table~\ref{tab:NCal_params}.)
Then, we computed the torque-corrected measured strain as
\begin{equation}
    h^{\rm meas \,\prime}_{\rm NCal} = \frac{F^{\rm meas \,\prime}_{\rm NCal}}{M (2 \pi f)^2 L_x}.
    \label{eqn:NCal_torque_corr}
\end{equation}


\section{Results}
\label{sec:results}


In this section, we compute and analyze the ratio between the measured and expected strain for each PCal and NCal injection. We then discuss to what degree these two calibration methods agree.

\subsection{Calibration ratios}
\label{sec:ratios}

\begin{figure}[hbt!]
    \centering
    \includegraphics[width=1.0\linewidth]{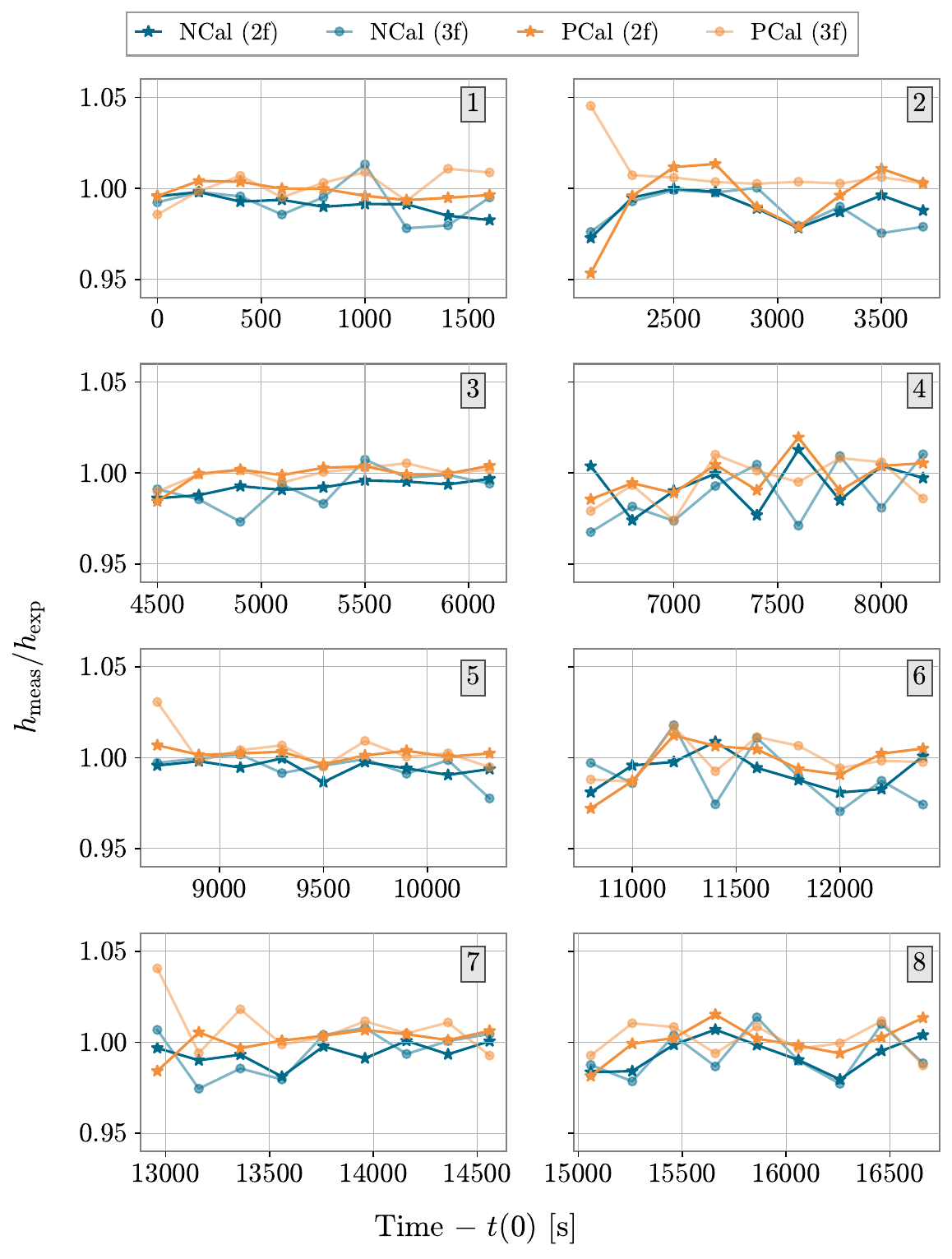}
    \caption{Ratio of measured to expected strain magnitude as a function of time at each $2f$ (star marker) and $3f$ (circle marker) harmonic for each NCal (blue) and PCal (orange) injection. The parameter $t(0)$ represents the start time of the experiment, GPS time = 1283182218~s. Each numbered panel spans 30~mins total and corresponds to one of the eight numbered rows shown in Table~\ref{tab:NCal_PCal_freqs}.}
    \label{fig:mags_v_time}
\end{figure}

\begin{figure}[hbt!]
    \centering
    \includegraphics[width=1.0\linewidth]{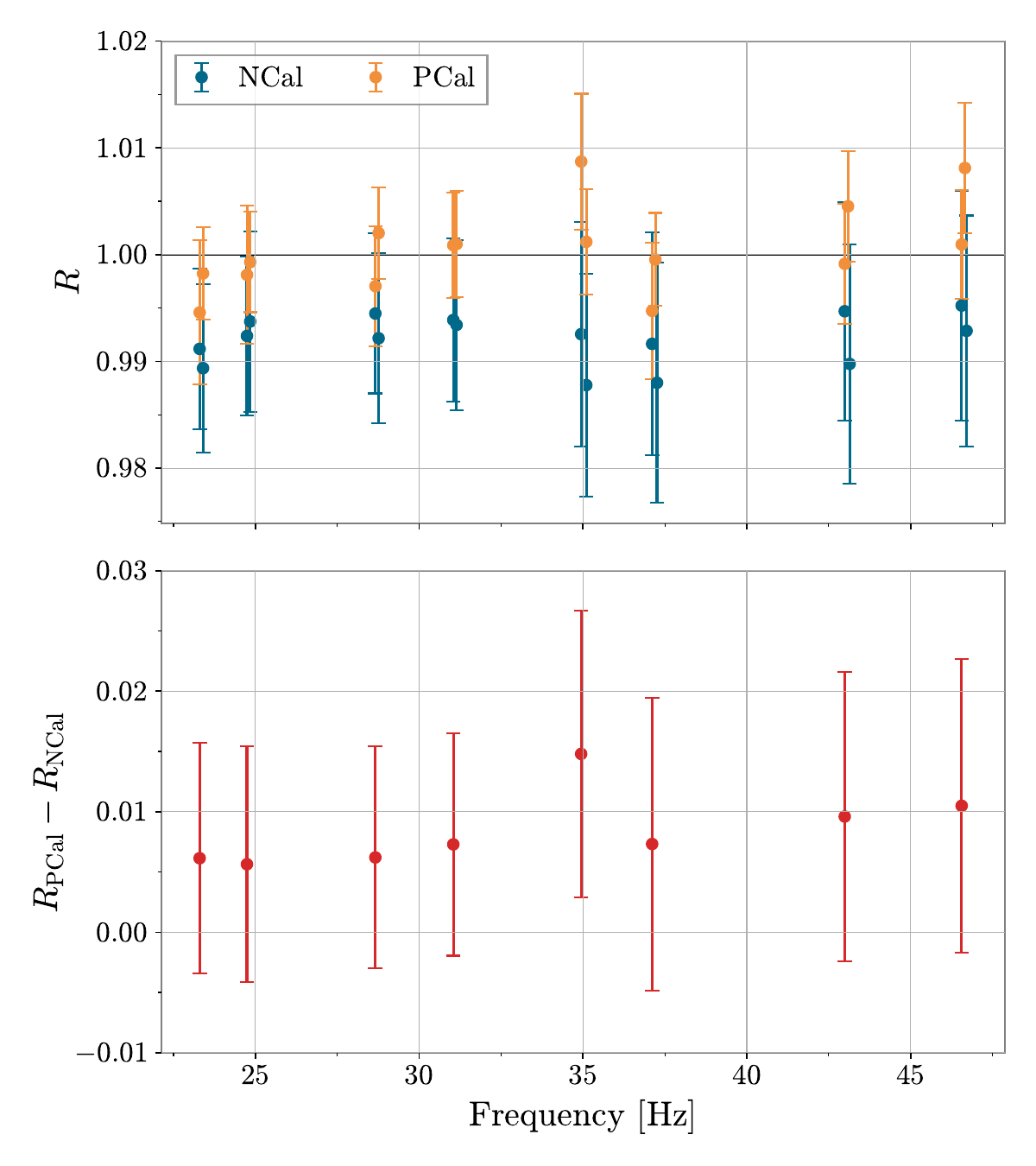}
    \caption{Top panel: Ratio of measured to expected strain magnitude as a function of frequency for the NCal (blue) and PCal (orange) injections. The error bars correspond to a $1 \sigma$ uncertainty on each ratio. Bottom panel: Residuals obtained by taking the average of each pair of NCal and PCal ratios separated by 0.1~Hz to obtain one value of $R_{\rm NCal}$ and $R_{\rm PCal}$, then subtracting $R_{\rm NCal}$ from $R_{\rm PCal}$ at each frequency. The bottom panel illustrates the deviation between the NCal and PCal measurements.}
    \label{fig:all_data}
\end{figure}

Using the strain amplitudes obtained in the previous section, we can find the ratios
\begin{equation}
    R_{\rm NCal} = \frac{h^{\rm meas \,\prime}_{\rm NCal}}{h^{\rm exp}_{\rm NCal}}, \quad
    R_{\rm PCal} = \frac{h^{\rm meas}_{\rm PCal}}{h^{\rm exp}_{\rm PCal}}
    \label{eqn:ratios}
\end{equation}
for each NCal and PCal injection.
Figure~\ref{fig:mags_v_time} shows each set of nine strain ratios across time (before they were averaged together) for NCal (blue) and PCal (orange) at $2f$ (star marker) and $3f$ (circle marker).
The averaged ratios are shown in the top panel of Fig.~\ref{fig:all_data} for NCal (blue) and PCal (orange) at their relevant frequencies.
The slight frequency dependence of the detector response, visible in the top panel of the figure, is both expected and well-documented. The response of the interferometer to an external force-induced differential displacement readout is not perfectly frequency independent; rather, it reflects the combined effects of the test mass and suspension dynamics, the optical response of the interferometer, and the feedback control systems. These frequency-dependent contributions are incorporated into the calibration model and lead to small deviations from a flat response across frequency. Similar frequency-dependence in the detector response can be seen in, e.g., the top panels of Fig.~15 of Ref.~\cite{Sun2020} and Fig.~16 of Ref.~\cite{Karki2016}.

Examining the data more closely, we notice that most data points lie within $1 \sigma$ of each other. However, for a given frequency, there is a systematic difference between $R_{\rm NCal}$ and $R_{\rm PCal}$ of about 0.5--1\%, with PCal consistently reporting slightly higher values. This is demonstrated in the bottom panel of Fig.~\ref{fig:all_data}, in which the residuals $R_{\rm PCal} - R_{\rm NCal}$ are plotted at each injection frequency. Nevertheless, we cannot say these two calibration methods are \textit{inconsistent}, given that the difference remains mostly within the margin of uncertainty.

The origin of the $\sim 1$\% difference between $R_{\rm NCal}$ and $R_{\rm PCal}$ remains unclear, although several possible explanations may account for this difference. The most likely culprit is simply that certain systematics in either the NCal or PCal systems have not yet been fully accounted for. However, the fact that we had only a limited dataset to work with may also be a factor.
In addition, several modifications have been made to the PCal system since these measurements were taken, so a new dataset with more recent calibration injections should be analyzed.


\subsection{Uncertainty}
\label{sec:unc}

While we defer the full error propagation to Appendix~\ref{appendix:error_prop}, we give an overview of the main sources of error in both the NCal and PCal systems here.

The dominant source of error in the NCal system is statistical uncertainty arising from measurement noise, largely due to the limited duration of the injections. Additionally, several sources of systematic uncertainty are present. One such source is indirect gravitational coupling within the suspension system, whereby the NCal rotor exerts forces not only on the test mass but also on the upper stages of the suspension system.
Non-gravitational couplings also contribute, including potential magnetic interactions between the rotor and test mass, as well as vibrations generated by the spinning rotor. These vibrations may couple into the strain measurement either through the seismic isolation system or via scattered light. Furthermore, as discussed in Sec.~\ref{sec:torque_corr}, the NCal rotor induces a torque on the test mass, introducing a rotational coupling in the strain measurement as a result of the offset of the main laser from the test mass center~\cite{Ross2021}.

Similar to the NCal system, statistical uncertainties arising from measurement noise are the dominant source of error in the PCal system.
Meanwhile, the largest single contributor to the systematic uncertainty is the absolute laser power calibration, set by the NIST-traceable calibration of the Gold Standard (GS) photodiode. An additional contribution comes from rotational coupling of the test mass due to offsets of the PCal beams from the center of the test mass, combined with the off-center position of the main interferometer beam, which introduces an additional geometric correction and associated uncertainty. Subdominant systematic contributions include the transfer calibration from the GS to the working photodiodes, temperature-dependent variations in photodiode responsivity, and uncertainties in the optical efficiency of the PCal system.  
All remaining contributions are negligible at the level of the total uncertainty~\cite{Bhattacharjee2020, Sun2021}.


\section{Conclusion}
\label{sec:conclusion}

In this study, we have analyzed and compared a series of direct-force injections made by the NCal and PCal systems at the LIGO Hanford observatory. The ratios of the measured to expected strain amplitudes for a series of frequencies between 20 and 50~Hz for each system have been found to agree to within approximately 1\%, although the PCal system reports slightly higher ratios across frequency.
While we can theorize about various overlooked systematics in either calibration system, future injections spanning longer time periods and wider frequency bands are needed to pinpoint the exact source of this discrepancy.

This study demonstrates the importance of having a second, independent, direct-force calibration technique to cross-check PCal, which has been the primary method of calibration in the LIGO detectors for many years. Resolving the discrepancy between NCal and PCal will represent a major step towards high fidelity calibration, and in the future, the two systems may even be used to jointly calibrate the LIGO detectors.

Although the observed systematic shift between NCal and PCal is not yet fully understood, it does not limit the overall calibration accuracy of LIGO. In current observing runs, the total calibration uncertainty is dominated not by the sub-percent uncertainties associated with the absolute displacement reference provided by PCal, but by uncertainties arising from imperfect modeling of the time- and frequency-dependent detector response, as well as statistical variations in the response measurements~\cite{Sun2020, Sun2021, Cahillane2017, Wade2025, GW150914_calibration}. These uncertainties contribute at the few-percent level across much of the sensitive frequency band.
However, as gravitational-wave detectors become more sensitive and observed signals achieve higher signal-to-noise ratios, calibration requirements for precision astrophysics will become increasingly stringent 
Next-generation detectors, such as Cosmic Explorer~\cite{Evans2021, Evans2023} and Einstein Telescope~\cite{ET_Abac_2026, Maggiore2020}, will require sub-percent calibration uncertainties to fully exploit their scientific potential, making systematic uncertainties in the absolute displacement reference increasingly relevant~\cite{Sun2021, Evans2021, Inoue2018}. Thus, as we draw ever closer to the future generation of gravitational-wave observatories, further studies investigating and mitigating errors and uncertainties in the absolute calibration methods are critical.

One factor expected to significantly improve the accuracy of the NCal system in future runs is the planned transition from a single rotor to a four-rotor array~\cite{Ross2023}.
In the current single-rotor system, calibration precision is limited by geometric uncertainties---in particular, the relative position of the rotor and test mass---which directly couple into the strain measurements. A multi-rotor configuration can mitigate this by producing a gravitational field that is less sensitive to metrology uncertainties.
The benefits of this approach have already been demonstrated by the multi-rotor NCal system installed in the Virgo detector~\cite{Aubin2024, Virgo_O4_cal}. Implementing a four-rotor array should therefore improve both the robustness and absolute accuracy of the NCal system, ultimately enabling more reliable measurements of the detector response.



\begin{acknowledgments}

The authors would like to thank Laurence Datrier, Corey Gray, Rahul Kumar, and Timesh Mistry for their role in facilitating the NCal and PCal injections and data collection.
The authors would also like to thank Beno\^{i}t Mours and Will Farr for their helpful comments during the internal review.
This material is based upon work supported by NSF's LIGO Laboratory, which is a major facility fully funded by the National Science Foundation.
LIGO was constructed by the California Institute of Technology and Massachusetts Institute of Technology with funding from the National Science Foundation, and operates under Cooperative Agreement PHY--1764464. Advanced LIGO was built under grant No. PHY--0823459.
The authors are grateful for computational resources provided by the LIGO Laboratory and supported by NSF Grants PHY--0757058, PHY--0823459.
This research was supported by the Australian Research Council Centre of Excellence for Gravitational Wave Discovery (OzGrav), Project Number CE230100016, Linkage Infrastructure, Equipment and Facilities, Project Numbers LE210100002 and LE260100008.
In addition, this research was supported by funding from the NSF under Awards PHY-1607385, PHY-1607391, PHY-1912380, and PHY-1912514.
L.S. is also supported by the Australian Research Council Discovery Early Career Researcher Award, Project Number DE240100206.
\end{acknowledgments}


\onecolumngrid
\appendix

\section{Error propagation}
\label{appendix:error_prop}

In this section, we give a detailed outline of how we estimated the total uncertainty in the measured and expected NCal and PCal strain magnitudes, shown as error bars in Fig.~\ref{fig:all_data}. The central values and uncertainties of several parameters used in these calculations can be found in Table~\ref{tab:NCal_params}.


\subsection{NCal measured strain}

We started with the standard error from Eq.~\eqref{eqn:meas_strain_unc}:
\begin{equation}
    \sigma_{h, \rm{NCal}}^{\rm meas} = \frac{s_{h, \rm NCal}^{\rm meas}}{\sqrt{N}}.
    \label{eqn:NCal_std_error}
\end{equation}
Next, we found the error on the measured force set by Eq.~\eqref{eqn:NCal_F_meas}:
\begin{equation}
    \sigma_{F, \rm{NCal}}^{\rm meas} = F^{\rm meas}_{\rm NCal} \sqrt{\Bigg(\frac{\sigma_{h, \rm NCal}^{\rm meas}}{h_{\rm NCal}^{\rm meas}}\Bigg)^2 + \bigg(\frac{\sigma_M}{M}\bigg)^2 + \bigg(\frac{2\sigma_f}{f}\bigg)^2 + \bigg(\frac{\sigma_{L, x}}{L_x}\bigg)^2}.
    \label{eqn:NCal_F_meas_unc}
\end{equation}
Following along with Sec.~\ref{sec:torque_corr}, we propagated the error through Eq.~\eqref{eqn:NCal_F_corr} as shown:
\begin{equation}
    \sigma_{F, \rm{NCal}}^{\rm meas\,\prime} = \frac{1}{F^{\rm meas \,\prime}_{\rm NCal}} \sqrt{
    \bigg(\Big[F^{\rm torq}_{\rm NCal} - F^{\rm meas}_{\rm NCal} \cos(\phi)\Big] \sigma_{F, \rm{NCal}}^{\rm torq}\bigg)^2 +
    \bigg(\Big[F^{\rm meas}_{\rm NCal} - F^{\rm torq}_{\rm NCal} \cos(\phi)\Big] \sigma_{F, \rm{NCal}}^{\rm meas}\bigg)^2
    }.
    \label{eqn:NCal_torque_corr_unc}
\end{equation}
Finally, we found the total uncertainty on the torque-corrected measured strain in Eq.~\eqref{eqn:NCal_torque_corr} to be
\begin{equation}
    \sigma_{h, \rm{NCal}}^{\rm meas\,\prime} = h^{\rm meas \,\prime}_{\rm NCal} \sqrt{\Bigg(\frac{\sigma_{F, \rm{NCal}}^{\rm meas\,\prime}}{F^{\rm meas \,\prime}_{\rm NCal}}\Bigg)^2 + \bigg(\frac{\sigma_M}{M}\bigg)^2 + \bigg(\frac{2\sigma_f}{f}\bigg)^2 + \bigg(\frac{\sigma_{L, x}}{L_x}\bigg)^2}.
    \label{eqn:NCal_meas_strain_unc}
\end{equation}

\subsection{NCal expected strain}

We calculated the total uncertainty on $h^{\rm exp}_{\rm NCal}$ in Eq.~\eqref{eqn:NCal_exp_strain} as shown:
\begin{equation}
    \sigma_{h, \rm{NCal}}^{\rm exp} = h^{\rm exp}_{\rm NCal} \sqrt{\Bigg(\frac{\sigma_{F, \rm{NCal}}^{\rm exp}}{F_{\rm NCal}^{\rm exp}}\Bigg)^2 + \bigg(\frac{\sigma_M}{M}\bigg)^2 + \bigg(\frac{2\sigma_f}{f}\bigg)^2 + \bigg(\frac{\sigma_{L, x}}{L_x}\bigg)^2}.
    \label{eqn:NCal_exp_strain_unc}
\end{equation}

\subsection{PCal measured strain}

The total uncertainty on $h^{\rm meas}_{\rm PCal}$ is simply the standard error from Eq.~\eqref{eqn:meas_strain_unc}, copied here for convenience:
\begin{equation}
    \sigma_{h, \rm PCal}^{\rm meas} = \frac{s_{h, \rm PCal}^{\rm meas}}{\sqrt{N}}.
\end{equation}

\subsection{PCal expected strain}

We again started with the standard error formula from Eq.~\eqref{eqn:PCal_exp_unc}:
\begin{equation}
    \sigma_{A, \rm PCal}^{\rm exp} = \frac{s_{A, \rm PCal}^{\rm exp}}{\sqrt{N}}.
\end{equation}
Next, we propagated the error through Eq.~\eqref{eqn:PCal_exp_strain}, wherein we converted the amplitude to strain units, to find the total statistical uncertainty:
\begin{equation}
    \sigma_{h, \rm PCal}^{\rm exp, stat} = h_{\rm PCal}^{\rm exp} \sqrt{\Bigg(\frac{\sigma_{A, \rm{PCal}}^{\rm exp}}{A^{\rm exp}_{\rm PCal}}\Bigg)^2 + \bigg(\frac{\sigma_{L, y}}{L_y}\bigg)^2}.
    \label{eqn:PCal_exp_strain_stat_unc}
\end{equation}
We also included the systematic uncertainty, expressed as
\begin{equation}
    \sigma_{h, \rm PCal}^{\rm exp, sys} = h_{\rm PCal}^{\rm exp} \, \sigma_{\rm PCal}^{\rm rel, sys},
    \label{eqn:PCal_exp_strain_sys_unc}
\end{equation}
where $\sigma_{\rm PCal}^{\rm rel, sys} = 0.0041$ across all of O3 (see Table 3 in Ref.~\cite{Sun2021}).
Finally, combining Eqs.~\eqref{eqn:PCal_exp_strain_stat_unc} and \eqref{eqn:PCal_exp_strain_sys_unc} in quadrature, we found the total uncertainty on the expected PCal strain: 
\begin{equation}
    \sigma_{h, \rm PCal}^{\rm exp} =\sqrt{\Big(\sigma_{h, \rm PCal}^{\rm exp, stat}\Big)^2 + \Big(\sigma_{h, \rm PCal}^{\rm exp, sys}\Big)^2}.
\end{equation}


%

\end{document}